# Fundamental Measurements in Economics and in the Theory of Consciousness
## (Manifestation of quantum-mechanical properties of economic objects in slit measurements)

I. G. Tuluzov[1] and S. I. Melnyk[2,*]

[1]*Kharkov Regional Centre for Investment, of.405, Tobolska str., 42a, 61072, Kharkov, Ukraine*

[2]*Kharkov National University of Radio Electronics, Lenin Ave., 14, 61161, Kharkov, Ukraine.*

A new constructivist approach to modeling in economics and theory of consciousness is proposed. The state of elementary object is defined as a set of its measurable consumer properties. A proprietor's refusal or consent for the offered transaction is considered as a result of elementary economic measurement. We were also able to obtain the classical interpretation of the quantum-mechanical law of addition of probabilities by introducing a number of new notions. The principle of "local equity" assumes the transaction completed (regardless of the result) of the states of transaction partners are not changed in connection with the reception of new information on proposed offers or adopted decisions (consent or refusal of the transaction). However it has no relation to the paradoxes of quantum theory connected with non-local interaction of entangled states. In the economic systems the mechanism of entangling has a classical interpretation, while the quantum-mechanical formalism of the description of states appears as a result of idealization of the selection mechanism in the proprietor's consciousness.

**Contents**

**Introduction**

In the first part of this paper [1] we have applied the methodology of the theory of selective measurements [2] to the analysis of economic systems. In particular, it has been proposed to describe the state of elementary economic system as a *set of its consumer properties determined as a possibility of exchange of certain elementary economic objects for others.* We assume that such determination of its state is complete and sufficient for describing the dynamics of economic systems. In other words, we restrict the description of the system's properties to its economic component, manifesting itself as a possibility of exchanging a certain economic object (showing its consumer properties in such acts of exchange) for another object with different consumer value.

In this model various product exchange procedures (transactions) can be considered as *economic measurements*. In this case, an offer of exchange of an elementary economic object for a different object corresponds to each elementary economic measurement. Along with this, in case if two different objects differ only in the quantity of conventional units of a certain product, they can be considered as a homogeneous class of objects and thus a corresponding discrete or continuous scale of measurement of consumer value can be developed. Scale mark corresponding to the quantity of product offered (or requested) for exchange characterizes its consumer properties quantitatively. Changing of its various consumer properties of the economic object corresponds to its exchanges for various products. Such measurement is determined similarly to the measurement of particle coordinate in physics using a screen with a slit. The upper edge of the slit corresponds to the seller's offer and the lower - to the buyer's offer. At the same time, the result of such measurement is not only a consent for a transaction (one of the participants accepts the other's offer), but also a refusal of it (corresponds to passing of the particle through the slit in the screen).

It is essential that in the general case such offer of transaction and subsequent refusal of it can change the consumer properties of the system (possibility of consent of refusal of further transactions), as they change the proprietor's notion of the value of his property. And this means that for the description of such measurement it is necessary to apply the quantum-mechanical formalism. Let us not, for the purpose of comparison, that in the classical model the state of object is changed only as a result of conclusion of a deal and remains unchanged in case of refusal of it.

Besides transaction-type economic measurements, we have discussed the technology-type measurements, in which the consumer properties of the discussed economic system are changed in a specific manner without the exchange procedure under the influence of external factors. Economic measurements of both types can be described using the symbols of generalized economic measurement $M(a',b')$. This symbol corresponds to the technology, in which the economic objects in state $a' \equiv \left(a' - \frac{\delta_a}{2}; a' + \frac{\delta_a}{2}\right)$ at the input and in state $b' \equiv \left(b' - \frac{\delta_b}{2}; b' + \frac{\delta_b}{2}\right)$ at the output are selected. The algebra of such generalized economic measurements has been developed. The symbols of addition and multiplication have been introduced and the required set of axioms with transparent economic meaning has been formulated for these measurements.

It has also been shown that in the general case a complete set (basis) of economically compatible values and corresponding measurements with independent results can be formed. The transformation function for linking descriptions of the generalized measurement in various complete sets has been obtained

$$M(c',d') = \sum_{a',b'} \langle a'|c'\rangle \langle d'|b'\rangle M(a',b') \qquad (1)$$

In the second part of the present paper, we are going to discuss the statistical meaning of the coefficients contained in (1). The economic analog of the slit experiment will be analyzed in detail and it will be shown that it can be used as an indicator of presence of quantum-mechanical properties of the observed economic systems.

### 1. Properties of economic measurements

#### 1.1. Statistical meaning of the transformation functions

Both in [2], and in the theory of economic measurements proposed by us, the coefficients of the transformation function $\langle a'|b'\rangle$ written before the symbols of measurements in (1) have only the meaning of the procedure of extraction of particles from one state into the other. They determine the equivalence relation of various formal combinations of the complete set of technologies. However, in order to be able to predict the numerical results of measurements we must link a certain **measurable** number to each such coefficient and formulate the rules of corresponding measurements.

Let us note for this purpose, that the only valid (measurable) result of the discussed elementary economic measurements is the consent of refusal to conclude the proposed transaction. All the rest of the results are either consequent from these elementary measurements, or do not relate to the sphere of economics. Therefore, the quantitative expression of the symbols $\langle a'|b'\rangle$ is to provide the calculation of these results. Let us note that the specific character of economic systems does not allow a precise prediction of the results of subsequent transactions on the basis of the results of the previous ones. In the general case, the result of the prediction is only the probability of consent (or refusal) for the proposed transaction for an economic object "prepared" in a specific way.

In physics, the basis for linking the symbols of the transformation function and the statistical regularities can be the properties of invariance of these symbols relative to their multiplication by an arbitrary phase multiplier [2]. As far as there is no physical specific character in this substantiation, and there is only the mathematical property, following from the axioms of the algebra of selective measurements, it can be completely and without changes attributed to economic symbols of measurement as well.

At the same time, notions relating to the probability and the set of alternatives in economic models are of a more transparent nature. This will further allow us to show that the quantum laws of addition of probabilities for alternatives occur as a simplified classical description of a more complete set of possible experimental results corresponding to various combinations of *incomplete economic measurements.* An attempt to describe the state of the observed system as a set of results of possible complete measurements results in the necessity of introducing the quantum-mechanical formalism. This simplified description will be discussed in detail in the process of the analysis of the slit experiment.

#### 1.2. Principle of local equity

The specific character of transactions, in contrast to technologies, is that the state of each of its participants is changed not as a result of physical (material) changes of property values, but as a result of receiving information about it along with various offers from the partner. Thus, we can say that the elementary economic measurement consisting of an offer of transaction and a positive or negative answer to it is itself the information technology responsible for changing of the state. At the same time, it turns out to be unimportant how quickly this event occurs and in what way this information will be received. Similarly, in the slit experiment with particles the thickness and material of the screen walls does not make any difference. For such "information" technologies the specific information meaning is associated with the notions of "input" and "output". Let us consider the procedure of trading common for such transaction, when the participants make offers to each other by turns. Let us suppose that the first offer is made by Alice. By accepting this offer, Bob "hits the lower semi-screen" at the input, and his further destiny is not considered by us. In case of his refusal (passing over the semi-screen), Alice receives information on his state at the input of the measurement, and this state of Bob has changed as a result of the offer made to him and his refusal of it. If Bob now makes a counter offer to Alice (specifies the price for which he agrees to work – upper semi-screen), it can be completely different from what he could have made if he had to make a "first move".

A good example of such situation is a scene of trading in Jack London's novel "Smoke and Shorty": "You are saying that if this land is not worth one hundred thousand dollars, it's not worth ten cents. At the same time, you are offering me five thousand. This means it's worth all one hundred thousand as well". A proprietor's state can change approximately in the same manner as a result of the received offers.

Thus, the result of refusal of consent for a pair of offers can depend on the sequence in which they were made. Asymmetrical situation can turn out to be "inequitable" for one of the participants. That is why in real trading the parties seek mutual consent for the final terms and conditions of the transaction. In other case, the bidding is stopped and the transaction is not concluded. However, even in this case we can consider that upon receiving the partner's offer, each of the participants "reserves his own opinion" and does not make any new offers. This means that his state no longer changes. In this case we can state that:

- the sequence of receiving offers from the participants of the transaction is no longer important and the results of measurements (their consent or refusal of the offers) do not depend the sequence of these offers. This assumption allows us drawing on the conventional scheme the upper and the lower edges of the slit in the same plane of the thin screen;
- the differences between the "input" and "output" of the transaction are eliminated and we can consider the state of particles passing through the slit identical both at the input and at the output.

Let us note that the analysis of slit experiments in physics mainly deals with the type of infinitely thin slits, and the channel-type measurement (Fig.1 (a)) is described as a continuous sequence of screens with finite width. Let us also stress that the property of "trade completeness" can be considered as a possibility of refusing a transaction immediately upon accepting offers. Such transaction should envisage confirmation of preliminary offers of each of the participants within a certain short period upon concluding an agreement in order that none of the participants could reproach the other one for providing unequal conditions. Therefore, we will further refer to the condition of trade completeness as the principle of local equity. In physics this condition is equivalent to the principle of local reversibility. It ensures repeatability of measurement results within a certain short period upon obtaining these results.

We are discussing this feature of real measurements (and corresponding transactions) in detail because we will further need to analyze a different type measurements, in which the results of measurements at the input and at the output do not match. In the economic context we will refer to such measurements as ***incomplete transactions.*** They are incomplete in the sense that at least one of the participants is willing to change his offer again or to change his previous answer to the offer of the other participant.

Thus, we are discussing the following hierarchy of slit interpretations of elementary economic measurements:

- The structure of fundamental measurements in economic models is based on elementary selective measurements consisting of one offer of transaction and one positive or negative answer. The equivalent of such measurement is a semi-infinite screen.
- An incomplete transaction consisting of two counter offers of each of its participants can be constructed from two such semi-screens. In this type of transactions the sequence of receiving offers is essential.
- In real transactions the trading procedure is possible, taking place until the counter offers of the participants become self-consistent. This condition corresponds to the principle of local equity and is represented by the slit in the thin infinite screen. In this case, the object's state at the input and at the output of the slit is the same.
- Actually, only the last type of locally equitable economic measurements corresponds to the slit measurements discussed in the analysis of the quantum-mechanical experiments in physics.

**2. Analysis of the slit experiment with subjects of economic relations**

*2.1. Classical and quantum screens in economic slit experiments*

So far, we have not been distinguishing economic objects according to the degree and character of their interference. Formally, any transaction between any economic objects can be considered as a selective measurement of properties of each of them by the other participant of the transaction. However, in physics the quantum-mechanical description is normally used for micro objects, while the classical description is used for macro objects. The reason for this is the elementary nature of the former type of objects and the complexity (in terms of degrees of freedom) of the latter. Therefore, we will also distinguish the transaction between two elementary economic objects in the aforesaid sense and the transaction between an elementary object and a macroscopic object. We have previously [1] defined the elementary economic object as an object, the consumer properties of which cannot be obtained from the consumer properties of its components.

In the first case both participants of the transaction are in symmetrical conditions. The state of both participants is changed both in case of consent of one of them from the offer of the second one, and in case of refusal of both of them. The state of such objects is entangled. At the same time, the entangled character is merely informational and economic. The exchange of products is not necessary for this. Even an elementary incomplete semi-screen type measurement is sufficient, when one of the participants offers his price and the other refuses. Let us suppose, for instance, that Alice considers her property more expensive than Bob's property, and vice-versa. In this case the entangled essence of their states is that the result of this transaction contains information about the state of each of its participants, but only in relation to the other participant's state. It cannot be used directly for predicting the results of other transactions of each of them. At the same time, as soon as the cost of Alice's property is defined in an additional experiment, it becomes possible to use the result of entangling for the assessment of Bob's property cost.

In case of interaction of an elementary economic object with a macroscopic object, the state of the latter is described by classical parameters averaged in a number of degrees of freedom. The influence of the microscopic object on these parameters is negligibly small. In other words, entangling of states occurs in this case as well, but its effect on the results of subsequent transactions is negligible. This effect is mathematically described in the decoherence, which is sufficiently well-studied in physics.

Thus, the screens used in the slit experiments, which have already become classical, are macroscopic objects. We are further going to discuss in detail only this specific type of measurements.

### 2.2. Slit experiment in economic modeling

The essence of such experiments has been described by us in detail in the first part of the present paper. Now we are going to discuss one of them in details - the experiment with double-width slit. We assume that it is one of the simplest experiments, in which the quantum-mechanical properties of the observed objects can be revealed. In this case we will consider the second participant of transaction (employer in our example) a macroscopic economic object, whose properties can be described in a classical way. Let us assume that a certain number of employees receive a job offer (transaction) which they can accept or refuse. We will distinguish two types of employers:

- **Upper semi-screen** with edge located at the height of $(a + \delta a)$ corresponds to the offer of transaction with a payment per working day at the specified price, in this case the **applicants who refuse this offer are screened**. Then the economic objects (applicants for the vacancy) "passing" under the semi-screen are those who accept the offer. They remain candidates for the vacancy and participate in the further selection. .

- **Lower semi-screen** with edge located at the height of $(a - \delta a)$ corresponds to the offer of transaction with a payment per working day at the specified price, in this case the **applicants who accept this offer are screened**. (they are hired for the job and do not participate in the further selection). Then the economic objects "passing" over the semi-screen are those who refuse the offer.

Thus, *the economic objects "passing" through the slit* $[a - \delta a; a + \delta a]$ *are those who agreed for the transaction at the price* $(a + \delta a)$*, but refused it at the price* $(a - \delta a)$*.* At the same time it is not known in advance whether the "passing" objects agree for the transaction at the price $(a)$, or refuse it. It is also unknown whether they accept the price $(a + \delta a)$ if they are offered a price $(a)$ instead of $(a - \delta a)$ or demand a higher price. Therefore, in the general case we cannot state that the number of objects "passing" through the slit $[a - \delta a; a + \delta a]$ is equal to the sum of those who pass through its upper or lower section.

This statement can be valid only for classical objects with the coordinate determined a priori (opinion of each of the applicants for the vacancy about the "fair" payment). And this opinion does not change depending on what offer he has made.

One of the main paradoxes of quantum statistics is the lack of classical interpretation of the amplitudes of probability. Regardless of what type of system we are discussing - physical or economic – the results of measurements are always the system's answers to various questions asked by the observer. Therefore, the initial results of statistical processing can be only ordinary probabilities calculated in frequency sense. At the same time, the laws of addition of these probabilities must be of classical nature. Nevertheless, in attempt of calculating the probabilities of various system states the necessity of applying new (quantum) statistics arises. We state that this necessity arises as an attempt to save the set of statistical results from subjectivism and to describe the state of the observed system as a state existing independently of the observers wish. The analysis of economic models in this sense is much more preferable as in this analysis the subject's state directly influences the state of the observed system. And there is no need to discuss the "freedom of choice" of the electron in similar experiments, as many authors seeking sensations are doing now in connection with the proving of the "free will theorem" [3]. Therefore, we will analyze in detail both the reasons of development of conceptions on the quantum probability in economic systems and grounds for considering such quantum description complete.

The generalized economic measurement can be both a transaction and a technology. Generally speaking, we can consider a more general type of measurements combining these two types. It can be useful in the analysis of the model of weak continuous fuzzy quantum measurements [4]. But now we are going to restrict ourselves to the analysis of them separately and will start with the analysis of the technologies. In our case the offers of payment in advance or upon completion of work correspond to the input and output of the generalized measurement associated with the technology. As this technology does not take effect immediately and as a result of its execution the state of the participants of transaction is changed, then in the general case it corresponds to a certain "black box" in which the input and the output are different. It is conventionally shown in Fig. 3(a) as a channel in the thick screen.

Let us also consider the physical analogy of this measurement. Let us assume that within the period of passing of the particle through the channel, it is effected by the constant gravity force, and as a result its vertical (measured) coordinate is changed and the states of particles at the input and at the output are different. The channel width determines the quantity of particles passing through it. It the channel is sufficiently narrow, the particles with low phase-space volume can pass through it (Fig. 1b). Now let us decrease the channel width twice while retaining its length. It is easy to see that not the half but only a quarter of particles, which could previously pass with obstructions, can now pass the half-width channel. This effect is purely classical. Its explanation and description does not require using quantum conceptions. In this case the formula

$$|a_+ + a_-\rangle\langle a_+ + a_-| = |a_+\rangle\langle a_+| + |a_+\rangle\langle a_-| + |a_-\rangle\langle a_+| + |a_-\rangle\langle a_-| \qquad (2)$$

means that the ensemble of particles passing through the wide slit can be considered as a mix of four ensembles, in each of which the particles passed through one of the four narrower channels (Figure 2).

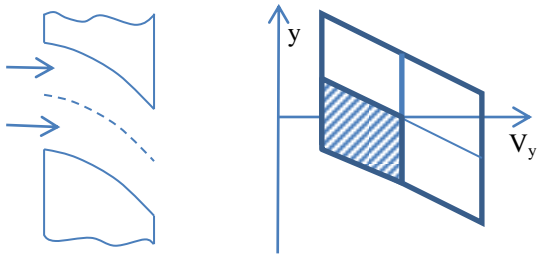

Figure 1(a) – narrow channel in the screen corresponding to the "technology"-type generalized measurement; 1(b) – phase volume corresponding to the ensemble of particles passing through the whole channel and through its lower section.

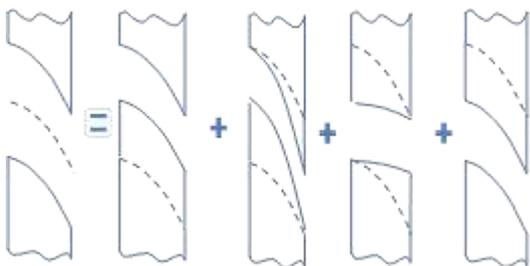

Figure 2. Classical ensemble interpretation of the formula of addition (distribution law) of symbols of the "technology"- type generalized measurements

Let us remind that the input of each of the channels corresponds to the conditions of advance payment, while the output corresponds to the payment upon completion of works. In case when the objects are workers applying for the job, we can assume that the initial axis velocity of the particle corresponds to the speed of natural decreasing of the consumer value of the hired worker due to the decrease of resources (labor force) or, on the contrary, corresponds to increasing of his consumer value due to professional training. Thus, the common classical probability theory is sufficient for the explanation of the effect of quadratic dependence of the number of economic objects passing through the classical channel. However, the "quantum mystics" occurs in the process of analysis of the transaction-type generalized measurements, in which no technological changes of the material consumer properties of the object occur, and the dependence remains quadratic. In this case, change of the state of the subject of transaction takes place practically instantly as a result of receiving information on the offer made to the him. The subject can, for instance, increase his price in his own opinion as a result of such offer and correspondingly increase the expected consumer value of his property (labor force). On the other hand, an offer of underestimated price can significantly decrease his demands.

Unlike the technology-type generalized measurement, in this case the point is not even that the changes take place almost instantly, but that it depends on the offer made. And the quantitative effect remains the same – only a quarter of particles can pass through the half-size slit.

### 2.3. "… the same, but «without a cat »"

Herbert Wells once gave a very simple description of the telegraph. «Imagine a giant cat, whose head is located in London and the tail in Liverpool. If someone in Liverpool steps on the cat's tail it will start mewing in London". After that, Wells explained the operation of wireless telegraph even simpler: "It's the same, but without a cat".

The same situation is observed in our case - we perfectly understand (and it does not contradict to our classical conceptions) why only a quarter portion of economic objects pass through the half-sized channel (technology). Thus all we have to do is to repeat Herbert Wells words that the transaction-type generalized economic measurement is even simpler - the same but without a channel.

In this case, the "input" and the "output" of the transaction are separated not by a time interval, during which the physical properties of the particle are changed, but by the information, which the economic objects (subjects of economic relations) receive in accordance with the position of the slit (estimated terms and conditions of transaction). In this case the formula (2) can be illustrated by Figure 3. Unlike Figure 1, in Figure 3 the channel is absent or it is very narrow, and the trajectories of objects can experience jumps in the screen plane. In physics such jumps mean superluminal velocity of particle and in the economic model it means instant (time interval is negligible) changing of the decision taken by the proprietor – consent or refusal of the offer of a specific price. In the discussed case, such offer is the offer of payment per one working day at the price (a). The worker receives two offers of employment at the price $(a + \delta a)$ и $(a - \delta a)$ between the input and the output of the economic slit. These offers can influence his state sufficiently to change his answer to the offer $(a)$.

At the same time, his consent or refusal of the offer at the input can influence his state at the output. In order not to complicate the situation, we will discuss only those experiments, in which this offer is made only at the input (before passing through the slit) or only at the output (upon passing through the slit). Such consideration is also convenient as it eliminates the questions of superluminal velocities of physical particle or ultra-fast changes of economic state, as these results relate to different measurements. At the same time, only one of them can be performed in relation to the particle (or economic object). The second measurement will be performed in relation to the changed state of the particle.

Thus, the result of one of the two possible measurements (at the input and at the output of the slit) are a sort of "hidden parameter", the existence of which is obvious, but the result itself is hidden from the observation. We will further analyze in detail the connection of these hidden parameters of economic models with physical "hidden parameters" of state of quantum objects, the possibility of existence of which is disproved experimentally.

The discussed jump trajectory illustrate in Figure 3 has only the sense of formal combination of the results of incompatible measurements. In order to stress the difference of these sets of results from the continuous trajectories analyzed both in classical and quantum dynamics we will further refer to them as the "*pseudo-trajectories*".

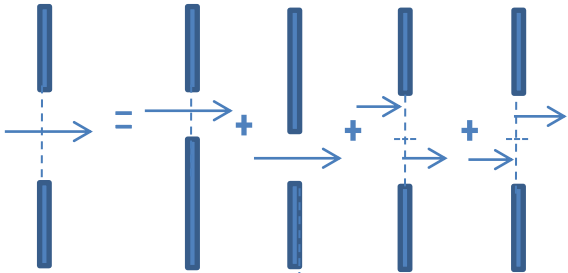

Figure 3. Scheme of the distributive law of addition of symbols of the transaction-type generalized measurements.

There is nothing either quantum-mechanical or mystical in these pseudo-trajectories. In case of considering them as a set of possible results of *alternative* measurements (both compatible and incompatible), it is sufficient to use classical statistics for the description of connection of different measurements.

Let us illustrate it on the example of the double-slit experiment (Figure 4(a), 4(b)). The slit in the first screen acts as a device responsible for preparation of particles in the same state. Figure 4(a) shows two alternative trajectories of motion of particles in Feynman meaning. The probability of reaching the slit of the third screen by the particle can only be calculated as a probability of superposition of these alternatives according to quantum-mechanical formulas, but not as a classical sum of probabilities of their mix.

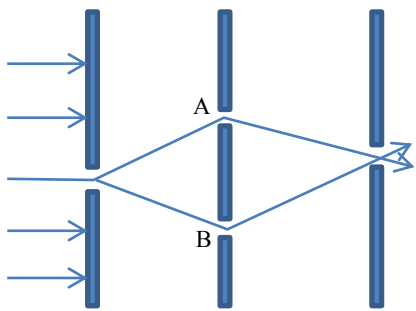

Figure 4(a)

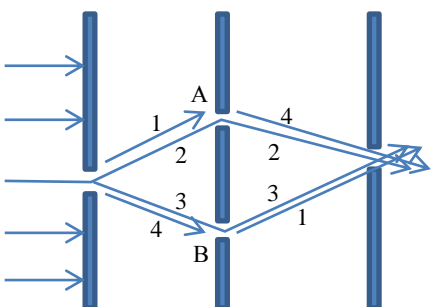

Figure 4(b)

In the process of analysis of the pseudo-trajectories, we must take account of the four possibilities (Figure 4(b)). Thus, for instance, the pseudo-trajectory (1) corresponds to those objects, which *would pass* though the upper section of the slit in case of measurement at the input, but would pass through the lower section in case of measurement at the output.

### 2.4. Complete classical description of state of measured economic objects in slit experiments

We have previously discussed that the offer $(a)$ made before the offers $(a + \delta a)$ or $(a - \delta a)$, corresponds to the input measurement, and the offer made after them corresponds to the measurement at the output of the completed transaction. At the same time, the offers $(a + \delta a)$ or $(a - \delta a)$ correspond both to the input and to the output. Thus, the notions "input" and "output" for the transaction-type generalized economic measurements are determined by the sequence of receiving of the offers. Therefore, the classical distributions of probabilities characterizing the state of the ensemble of measured economic objects should be analyzed only in relation to a fixed sequence of offers.

The simplest elementary economic measurement is a single offer of transaction. In our case it is one of the three offers $(a); (a + \delta a)$ or $(a - \delta a)$. At the same time we can find out the result of only one of them, after that the object's state will be changed. However, we can also assume that in case of performing the other two elementary measurements **we would** obtain certain definite results. Speaking only of real measurements, the sequence of their results for a set of prepared objects can be represented in the form of a three-row table, with one known result of measurement and two unknown results in each of its columns (Table 1).

**Table 1**. The table of possible results of elementary measurements can be filled in for a real experiment only by one third, as after the first measurement the object's state will be changed.

|  | 1 | 2 | 3 | 4 | 5 | … |
|---|---|---|---|---|---|---|
| $(a + \delta a)$ | + | ? | ? | ? | ? | - |
| $(a)$ | ? | - | ? | + | - | ? |
| $(a - \delta a)$ | ? | ? | - | ? | ? | ? |

Such table can be perfectly used as a model of hidden parameters (taking into account both the filled-in and not filled-in results), but the information contained in it is insufficient for a complete description of the state of objects, as there is no account of the influence of the first measurement on the results of the second one. At the same time, there are at least 2 offers in the discussed slit experiment. Therefore, a more complete description is provided by a set of six tables with the results of answers for the two sequenced offers filled in two rows. We are hereby representing an example of two such tables (Table 2) for the offers $(a + \delta a)$ and $(a)$ received in a different sequence.

The results of answers for the offers of transaction can depend on the sequence of offers. For a complete description of states of the economic object, it is necessary to set his answers for all possible pairs of offers with account of their sequence. The Table 2 illustrates tables for two such pairs of six possible.

**Table 2.**

|   |            | 1 | 2 | 3 | 4 | 5 | … |
|---|------------|---|---|---|---|---|---|
| 2 | $(a+\delta a)$ | + | + | + | + | - | … |
| 1 | $(a)$        | - | - | + | + | - | … |

|   |            | 1 | 2 | 3 | 4 | 5 | … |
|---|------------|---|---|---|---|---|---|
| 1 | $(a+\delta a)$ | + | + | - | + | - | … |
| 2 | $(a)$        | + | - | - | + | - | … |

………………………………………………..

Let us note that in case of changing the sequence of receiving of the two offers the answers to both of them can be different. However, there is a number of logically justified obvious limitations for possible ways of filling in these two tables. It is the prohibition for consent (+) for the offer $(a)$ on case of refusal (-) for the offer $(a+\delta a)$ and other similar offers. Based on these two tables we can select the set of those objects, which will pass through the slit $(a+)$. For this purpose we need to select those columns, for which a combination of symbols (+ -) is filled in both tables. In Table 2, it is the object 2.

Only for this type of objects the principle of local equity is valid and the transaction can be considered completed (but not performed, as the object passes through the slit). For the rest of the objects the conditions of passing through the slit are not valid either at the "input" or at the "output". At the same time the notions of input and output in this case are conventional (unlike the technologies), as the subject of our interest are the results for which the sequence of receiving offers is not important.

The three discussed variants of price offer with account of sequence give six possible pairs of offers. Therefore, for a complete description of all possible situations in the discussed slit experiment we will need two additional pairs of similar filled-in double-row tables for two more pairs of offers. These six tables act as hidden parameters, not the Table 1 alone. At the same time, in the real experiment we "open" the values of possible answers only in one of the six tables for each measured object. Besides, the first cell will be automatically "opened" in another table, in which the same offer is filled-in in the first place. Let us note that the derivation of Bell's inequalities or their economic analogs [9] only relates to Table 1, not to the set of table 2. In the proposed description of possible results of transactions and answers to offers, there is still no "quantum mechanics". This description is completely classical. Its difference from the theory of "hidden parameters" in physical models is that we are clearly introducing a dependency of the answer to an offer on what offer has been made before it.

We can also consider a model in which the answer to various offers depends not only on the previous offer, but also depends to a various degree on all the preceding offers. Such generalization is possible for the quantum mechanical formalism and is described in the theory of weak continuous quantum measurements. The mathematical apparatus of such theory has been used by us in our paper [5], where it has been used to obtain the quantum-mechanical generalization of the Black-Scholes formula. In this paper we pursue different aims and therefore we will restrict ourselves to the discussion of a more "trivial" case of fuzzy discrete measurements. At the same time, we can assume that the state of an economic object is completely determined only by the last results of measurement of the complete set of parameters of state.

Unlike physics, the problem of non-locality of interaction does not arise in the discussed model. In this model the interaction is local, as the answers given to the two questions by the same subject of transaction originate in the consciousness of the same subject. At the same time, the mechanism of interrelation of the two answers can be any, including simple classical scenarios. Following the terminology proposed by R. Penrose [6], in such description we are trying to solve a Z-paradox connected with the inconsistency of the logical interpretation of quantum measurements, but do not claim to solve the X-paradox relating to the explanation of the mechanism of non-locality of their interconnection. Let us also note that the observability of a pair of rows from six tables has an obvious sense only in economic models. It is an incomplete transaction, in which only two answers for the first two offers made in a specific sequence are taken into account. The question of an analog of incomplete transactions in physics and the possibility of such measurements remains open. Nevertheless, both in economics and in physics only completed transactions (measurements) are usually taken into account in the process of calculation of the probabilities. In the discussed example it is the passing of object (particle) through the upper section of the slit, lower section or through the completely open slit. In other words, in economic statistics the measurable parameters are the probabilities of completed transactions, in which the principle of local equity is observed, while the sequence of receiving offers does not influence the result of the transaction. For instance, the measurable parameter is the probability $^{a+\delta}_{a}w$ of passing through the upper section of the slit, which is equal to the percentage of objects, for which in both tables 2 the values $(\pm)$ are filled-in. Formally this condition can be written as
$$^{a+\delta}_{a}w = {^{a+\delta}_{a}}\rho(\pm:\pm).$$

### 2.5. Density matrix in economic measurements

One of the forms of notation of state of quantum objects is the formalism of the density matrix. In this case it becomes possible to describe not only pure states of the ensemble of observed economic objects, but also mixed states. The former can be obtained as a result of *the same* transaction, offered to an ensemble of proprietors prepared in the same way. It follows from the algebra of selective measurements that in the most general case the state of objects that have refused this transaction corresponds to the particles which have passed through the set of slits in physical models. This state can be described as the linear superposition of basis states $|i\rangle$ of the complete set of compatible variables $|x\rangle = \sum_{i=1}^{n} \alpha_i |i\rangle$.

Such measurement corresponds to the density matrix of pure state, which can be written as

$$\rho_{ij} = (\sum_{i=1}^{n} \alpha_i |i\rangle)(\sum_{i=1}^{n} \alpha_i^* \langle i|) \qquad (3)$$

In the general case, the coefficients $\alpha_i$ can be complex. However, in the discussed mental slit experiment they have clear meaning and correspond to the width of a certain slit in an infinite screen. Let us note that it follows not from the subsequent probabilistic interpretation of these coefficients, but only from the definition of the symbol of addition "+" of the selective measurements described by us previously. It is obvious that a narrow slit of double width can be considered as logical (in the aforesaid sense) sum of two single slits, located next to each other and described by approximately the same symbol of measurement. Accordingly, the n-times reduction of width will result in the coefficient 1/n before the corresponding basis vector. We will further consider these coefficients valid, assuming the possibility of generalization of the proposed illustrative model. In this case $\alpha_i \alpha_j^* = \alpha_j \alpha_i^* = \alpha_i \alpha_j$ and the density matrix turns out to be symmetric. The mixed state can be obtained as a result of a measurement, in which various ensembles of objects are offered various transactions, and then the refusing proprietors are mixed in a certain proportion. Mathematically it can be written as a weighted arithmetical total of density matrixes of each of the pure states.

At the same time, the density matrix corresponding to the pure state can be represented by a sum of matrixes with only one non-zero coefficient in each of them. It can be written before the matrix in the form of multiplier:

$$(\sum_{i=1}^{n} \alpha_i |i\rangle)(\sum_{i=1}^{n} \alpha_i^* \langle i|) = \sum_{i,j=1}^{n} \alpha_i \alpha_j^* |i\rangle\langle j| \qquad (4)$$

Formally, such notation can be considered as a mix of states obtained using the generalized selective measurements $|i\rangle\langle j|$, in which the objects are in state $\langle j|$ at the input, and in state $|i\rangle$ at the output. In the technology-type generalized economic measurements it is possible as the state of object is changed in accordance with the purpose of the technological process. In case of considering the transaction-type generalized economic measurements, the input and the output are not separated by any event. It would be wrong to expect that a proprietor "entering" through the upper section of the slit in our example and refusing the offer (a) will immediately "exit" through the lower section, and will have to accept the same offer in order to do so. However, we can easily assume that the state of object is changed as a result of receiving another offer with subsequent change of the proprietor's opinion on the cost value.

Let us return to the illustrative slit experiment with double-width slit. Let us select for consideration only those objects, which pass through the slit. It means that in the third pair of tables only «+;-» symbols are filled-in, regardless of the sequence of receiving offers $(a + \delta a)$; $(a - \delta a)$. We will assume that the result of influence of these offers on the answer (a) does not depend of their sequence. Then the input measurement of object corresponds to the case when the first offer is (a) and the subsequent offer is $(a + \delta a)$; $(a - \delta a)$. If the offer (a) is received after them, this case corresponds to the changing of the object's position (upper or lower section of the slit) at the output.

Then the combinations of symbols in all 6 tables in case of such selection will differ only in the values of the answer (a) at the input and at the output. Accordingly, the set of objects passing through the wide slit will be separated into 4 subsets, each corresponding to the summand in the formula 2 and to the corresponding coefficient in the density matrix. In case when the answers for the offer (a) at the input and at the output are independent, the probability of the combination input-output in equal to the classical product of these probabilities. At the same time, the coefficient of the density matrix satisfy the condition

$$\alpha_{11} \cdot \alpha_{22} = \alpha_{12} \cdot \alpha_{21} \qquad (5)$$

and such state is called pure. It can be written as

$$|x\rangle = \alpha_1 |a_+\rangle + \alpha_2 |a_-\rangle \qquad (6)$$

where $\alpha_1$ and $\alpha_2$ are the probability amplitudes of consent or refusal of the offer (a), respectively. In case of the narrow slit these are proportional to the width of the upper $(a; a + \delta a)$ and lower $(a - \delta a; a)$ section of the slit, respectively. Unlike the case of the technology-type generalized measurements, we cannot consider that the object from the subset $|a_+\rangle\langle a_-|$ "enters" through the upper section of the slit and "exits" through the lower one. These objects pass through the double slit, but will not pass through either its upper or lower section separately. The values $|a_+\rangle$ и $\langle a_-|$ correspond to the "pseudo-trajectory" – the values of coordinate, which **could be obtained** in case of input and output measurements performed separately. These measurements of economic objects are incomplete transaction and there is no pure state corresponding to them in the sense of formula (6). For such subsets, there are no means of preparing such states in physical analogs, as the meaning of the analog of incomplete measurement is not clear insofar. For economic objects, on the other hand, not only the "locally equitable" results of completed transactions, but also the intermediate results of incomplete transactions can be followed.

Therefore, in case of the description of this kind of sets in economics we can restrict ourselves to the frameworks of the classical theory of probabilities retaining the prefix "*if*" for all such results. Transition to the quantum-mechanical description and quantum amplitudes of probabilities is the "price" for simplification of the mathematical apparatus and refusal of the conventional formulation of the results of description of state.

In the general case, pure quantum-mechanical states of marker (spin) variables in physics can be described by the density matrix, in which the non-diagonal elements are complex. Detailed analysis of economic analogs of such states requires a separate consideration. So far, we can only assume that they occur in case of introducing prices for options and futures into the terms and conditions of the transaction, which are the analog of the result of measurement of particle momentum in physics.

Finally, let us note that in case of narrow slit, its upper and lower sections are practically equivalent, and the probability of receiving consent or refusal of the offer (a) can be considered equal both at the input and at the output. This statement relates to those objects, which pass through the double-width slit. Therefore, for such experiments the quantity of passing objects is proportional to the square of this width. By gradually widening the slit, the transition from the quadratic law to the linear law (corresponding to the classical sum of subsets of objects, for which the uncertainty of the coordinate is significantly smaller than the slit width) can be analyzed. The boundary of such transition from quantum to classical description can be used for estimation of the effective "wavelength" of the prepared set of economic objects.

The logical interpretation of other more complex quantum effects, such as interference and diffraction, will be possible at the next stage of construction of the theory, in the process of analysis of the geometry of space of states and the dynamics of fundamental economic states in serial selective measurements separated by a sufficient period of time.

**3. Proceeding from slit experiments to models of dynamics of economic systems**

In the economic interpretation of completed measurement it is assumed that the principle of local equity results in the stable state of the participants of transaction within a negligible time interval. In this state they do not refuse their previously made decisions and the transaction is completed either by accepting one of the offers, or by a definitive reject. In the aforesaid slit experiments it corresponds to an infinitely thin screen and, accordingly, to instant collapse of the quantum state. In this kind of approximation we are not interested in the mechanism of agreement of decisions (trading) and all the occurring events relate to the same time moment.

Following Schwinger's ideology, we can consider the evolution of state as a generalized selective measurement, which changes the object's state in time moment $t$ into another state in the next time moment $t + \delta t$. The aforesaid technology-type generalized economic measurements have such properties. They connect two states of an economic object in different time moments. Changing of the object's state in the interval between these moments does not envisage decision-making by the proprietor and is determined by external (in relation to him) factors. For instance, in the discussed example all workers get tired by the end of the working day and their price coordinate is decreased (Figure 1-2). This decrease does not depend on their choice, though it can be different for different workers in different initial state. The reasons of such effect can also be different. Its description, similarly to physics, can be phenomenological or can be based on specific properties of symmetry. Discussion of these issues is beyond the scope of the theory of measurements. However, if we know how this effect changes the state of object $|a_1\rangle$, and the state $|a_2\rangle$, the theory of generalized economic measurements allows calculating the changing $|x\rangle = \alpha_1|a_1\rangle + \alpha_2|a_2\rangle$ as a result of this effect.

As a result, it turns out that the whole dynamics of economic systems can be represented as alternate effects of the transaction-type and technology-type generalized measurements. It is an analog of effect of projection operators and unitary operators in modeling of the dynamics of physical systems. The former correspond to the procedure of changing of quantum sate and the latter – to the reversible evolution of the enclosed quantum system.

Such simplified model of alternate effects in the process of transition to the description of the continuously observed quantum system results in the "Zeno quantum effect". For adequate description of such observations it is necessary to use the generalized approach of the theory of continuous fuzzy quantum measurements. In the economic systems we can also consider dynamics as a continuous fuzzy measurement. In particular, we can adapt the known derivation of the Black-Scholes formula with account of quantum properties of state [5]. Similar approach can be used for modeling of the dynamics other economic objects.

**Conclusion**

Summarizing the discussion of the virtual slit experiment of economic measurement, we can state that we have obtained the economic analog of the quantum-mechanical description of the procedure of measurement of the generalized coordinate in physics. At the same time the description of the set of states of economic systems is based on the general principle, in which the system's state is determined only by the results of measurement. It has been proposed in the framework of the general concept of the generalized economic measurement to consider two types of measurements. The first transaction-type measurements correspond to the selection of objects passing through the slit or a set of slits in the thin screen. The second technology-type measurements correspond to the selection of objects passing through the channel with different (in the general case) input and output in the thick screen. It has been shown that the states of objects in the transaction-type measurements can be adequately described only with account of possible influence of the received offers of transactions. Unlike the classical case, the object's state changes not only as a result of concluding a transaction, which is obvious, but also as a result of refusal. In this case, the state of such objects is influenced by the information effect. The probabilistic description of such economic states can be represented as a description of the classical set of virtual events containing the proprietor's answers to various ordered pairs of offers. At the same time, in a real experiment answers to only one such pair can be received, after that the objects state should be considered as a changed state. The quantum-mechanical description of states and the corresponding laws of addition of amplitudes of probabilities for alternative events can be obtained as a simplified description linking the probabilities of only real performed experiments.

The further development of the method of generalized selective measurements applied to economic systems allows expecting a rigorous derivation of the laws of motion of economic systems, both isolated from the environment and those being in the continuous fuzzy observation mode.

______________________________________

[*]Electronic address: smelnyk@yandex.ru